\documentclass[aps,prd,preprint,floatfix,nofootinbib,superscriptaddress]{revtex4}
\usepackage{graphicx,subfigure,epsfig,epsf}

\def\poutsla{/\!\!\!p_{out}}
\def\pinsla{/\!\!\!p_{in}}

\newcommand{\beq}{\begin{equation}}
\newcommand{\eeq}{\end{equation}}
\newcommand{\bea}{\begin{eqnarray}}
\newcommand{\eea}{\end{eqnarray}}

\newcommand{\Lda}{\Lambda}
\newcommand{\g}{\gamma}
\def\Re{{\cal R \mskip-4mu \lower.1ex \hbox{\it e}\,}}
\def\Im{{\cal I \mskip-5mu \lower.1ex \hbox{\it m}\,}}

\def\etal{{\it et al.}}

\def\tev{\,{\ifmmode\mathrm {TeV}\else TeV\fi}}
\def\gev{\,{\ifmmode\mathrm {GeV}\else GeV\fi}}
\def\mev{\,{\ifmmode\mathrm {MeV}\else MeV\fi}}
\def\to{\rightarrow}

\begin{document}


\def\issue(#1,#2,#3){#1 (#3) #2} 
\def\APP(#1,#2,#3){Acta Phys.\ Polon.\ \issue(#1,#2,#3)}
\def\ARNPS(#1,#2,#3){Ann.\ Rev.\ Nucl.\ Part.\ Sci.\ \issue(#1,#2,#3)}
\def\CPC(#1,#2,#3){Comp.\ Phys.\ Comm.\ \issue(#1,#2,#3)}
\def\CIP(#1,#2,#3){Comput.\ Phys.\ \issue(#1,#2,#3)}
\def\EPJC(#1,#2,#3){Eur.\ Phys.\ J.\ C\ \issue(#1,#2,#3)}
\def\EPJD(#1,#2,#3){Eur.\ Phys.\ J. Direct\ C\ \issue(#1,#2,#3)}
\def\IEEETNS(#1,#2,#3){IEEE Trans.\ Nucl.\ Sci.\ \issue(#1,#2,#3)}
\def\IJMP(#1,#2,#3){Int.\ J.\ Mod.\ Phys. \issue(#1,#2,#3)}
\def\JHEP(#1,#2,#3){J.\ High Energy Physics \issue(#1,#2,#3)}
\def\JPG(#1,#2,#3){J.\ Phys.\ G \issue(#1,#2,#3)}
\def\MPL(#1,#2,#3){Mod.\ Phys.\ Lett.\ \issue(#1,#2,#3)}
\def\NP(#1,#2,#3){Nucl.\ Phys.\ \issue(#1,#2,#3)}
\def\NIM(#1,#2,#3){Nucl.\ Instrum.\ Meth.\ \issue(#1,#2,#3)}
\def\PL(#1,#2,#3){Phys.\ Lett.\ \issue(#1,#2,#3)}
\def\PRD(#1,#2,#3){Phys.\ Rev.\ D \issue(#1,#2,#3)}
\def\PRL(#1,#2,#3){Phys.\ Rev.\ Lett.\ \issue(#1,#2,#3)}
\def\PTP(#1,#2,#3){Progs.\ Theo.\ Phys. \ \issue(#1,#2,#3)}
\def\RMP(#1,#2,#3){Rev.\ Mod.\ Phys.\ \issue(#1,#2,#3)}
\def\SJNP(#1,#2,#3){Sov.\ J. Nucl.\ Phys.\ \issue(#1,#2,#3)}


\bibliographystyle{revtex}

\title{Neutral Higgs boson pair production at the LC in the Noncommutative Standard Model } 




\author{Prasanta~Kumar~Das}
\email[]{Author(corresponding) : pdas@bits-goa.ac.in, pdasMaparna@gmail.com}
\author{Abhishodh Prakash}
\email[]{abhishodh@gmail.com}
\author{Anupam~Mitra}
\email[]{anupam.mitra@gmail.com}

\affiliation{Birla Institute of Technology and Science-Pilani, K K Birla Goa campus, NH-17B, Zuarinagar, Goa-403726, India}


\date{\today}

\begin{abstract} 
We study the Higgs boson pair production through $e^+~e^-$ collision in the noncommutative(NC) extension of the standard 
model using the Seiberg-Witten maps of this to the first order of the noncommutative parameter $\Theta_{\mu \nu}$. This process is forbidden in the standard model with background space-time being commutative. We find that the cross section of the pair production of Higgs boson (of intermediate and heavy mass) at the future Linear Collider(LC) can be quite significant for the NC scale $\Lambda$ lying in the range $0.5 - 1.0$ TeV.  Finally, using the direct experimental(LEP II, Tevatron and global electro-weak fit) bound on Higgs mass, we obtain bounds on the NC scale as  $665$ GeV $\le \Lambda \le 998$ GeV. 
\end{abstract}

\maketitle


\section{Introduction }
Inspite of its enormous experimental success the standard model(SM) of particle physics still awaits the discovery of the 
Higgs boson. After the Large Electron Positron (LEP) collider has set a lower limit of about 114.4 GeV on its mass \cite{TJ}, 
the responsibility of finding the Higgs now rests mostly on the Large Hadron Collider (LHC) at CERN which has started its operation. At the same time, puzzles such as the naturalness problem make a strong case for physics beyond the standard model(SM), just around or above the mass scale where the Higgs boson is expected to be found. It is therefore of supreme interest to see if 
the collider signals of the Higgs boson contain some imprint of new physics. This necessitates detailed quantitaive exploration 
of a variety of phenomenona linked to the production and decays of the Higgs. 

In this paper, we have studied pair production of the Higgs boson in the intermediate and heavy mass range at the Linear Collider(LC) as a possible channel for uncovering new physics effects.  In particular, we show that such pair production which is forbidden in the SM with commutative space-time(we will call this as CSM in abbreviation) receives a large contribution in the noncommutative(NC) extension of the SM. 

As has been mentioned above, the large hierarchy between the electroweak scale $M_{W}$ and the Planck scale $M_{\it Pl}$ is  
somewhat puzzling. Though theories like supersymmetry and technicolour, each with its own phenomenological implications
and constraints,  have been proposed as a resolution of this problem, the idea of extra spatial dimensions with the scale of 
gravity being as low as TeV, has drawn a lot of interest among the physics community recently \cite{VR}. In some brane-world models \cite{ADD98} where this TeV scale gravity is realised, one can principly expect to see some stringy effects in the upcoming TeV colliders and in addition the signature of space-time noncommutavity. Interests in the noncommutative(NC) field theory arose from the pioneering work by Snyder \cite{Snyder47} and has been revived recently due to developments connected 
to string theories in which the noncommutativity of space-time is an important characteristic of D-Brane dynamics at low energy 
limit\cite{Connes98,Douglas98,SW99}. Although Douglas \etal \cite{Douglas98} in their pioneering work have shown that noncommutative field theory is a well-defined quantum field theory, the question that remains is whether the string theory prediction and the noncommutative effect can be seen at the energy scale attainable in present or near future experiments instead of the $4$-$d$ Planck scale $M_{pl}$. A notable work by Witten \etal \cite{Witten96} suggests that one can see some stringy effects by lowering down the threshold value of commutativity to \tev, a scale which is not so far from present or future collider scale. 

~ What is space-time noncommutavity? It means space and time no longer commute with each other and as a result one cannot measure the space and time coordinates simultaneously with the same accuracy. Writing the space-time coordinates as operators  we find 
\beq 
[\hat{X}_\mu,\hat{X}_\nu]=i\Theta_{\mu\nu}
\label{NCSTh}
\eeq
where the matrix $\Theta_{\mu\nu}$ is real and antisymmetric. The NC parameter $\Theta_{\mu\nu}$ has dimension of area and reflects the extent to which the space-time coordinates are noncommutative i.e. fuzzy. Furthermore, introducing a NC scale  $\Lda$, we rewrite Eq. \ref{NCSTh} as 
\beq 
[\hat{X}_\mu,\hat{X}_\nu]=\frac{i}{\Lda^2} c_{\mu\nu}
\label{NCST}
\eeq
 where $\Theta_{\mu\nu}(=c_{\mu \nu}/\Lda)$ and $c_{\mu\nu}$ has the same properties as $\Theta_{\mu\nu}$. To study an ordinary field theory in such a noncommutative fuzzy space, one replaces all ordinary products among the field variables with Moyal-Weyl(MW) 
\cite{Douglas} $\star$ products defined by
\begin{equation}
(f\star
g)(x)=exp\left(\frac{1}{2}\Theta_{\mu\nu}\partial_{x^\mu}\partial_{y^\nu}\right)f(x)g(y)|_{y=x}.
\label{StarP}
\end{equation}
Using this we can get the NCQED Lagrangian as
\begin{equation} \label{ncQED}
{\cal L}=\frac{1}{2}i(\bar{\psi}\star \gamma^\mu D_\mu\psi
-(D_\mu\bar{\psi})\star \gamma^\mu \psi)- m\bar{\psi}\star
\psi-\frac{1}{4}F_{\mu\nu}\star F^{\mu\nu} \label{NCL},
\end{equation}
which are invariant under the following transformations 
\bea
\psi(x,\Theta) \to \psi'(x,\Theta) &=& U \star \psi(x,\Theta), \\
A_{\mu}(x,\Theta) \to A_{\mu}'(x,\Theta) &=& U \star A_{\mu}(x,\Theta) \star U^{-1} + \frac{i}{e} U \star \partial_\mu U^{-1},
\eea
where $U = (e^{i \Lambda})_\star$. In the NCQED lagrangian (Eq.\ref{ncQED})
$D_\mu\psi=\partial_\mu\psi-ieA_\mu\star\psi$,$~~(D_\mu\bar{\psi})=\partial_\mu\bar{\psi}+ie\bar{\psi}\star
A_\mu$, $~~ F_{\mu\nu}=\partial_{\mu} A_{\nu}-\partial_{\nu}
A_{\mu}-ie(A_{\mu}\star A_{\nu}-A_{\nu}\star A_{\mu})$. 

 The alternative is the Seiberg-Witten(SW)\cite{SW99,Douglas98,Connes98,Jurco} 
approach in which both the gauge parameter $\Lambda$ and the gauge field $A^\mu$
is expanded as 
\bea \label{swps}
\Lambda_\alpha (x,\Theta) &=& \alpha(x) + \Theta^{\mu\nu} \Lambda^{(1)}_{\mu\nu}(x;\alpha) + \Theta^{\mu\nu} \Theta^{\eta\sigma} \Lambda^{(2)}_{\mu\nu\eta\sigma}(x;\alpha) + \cdot \cdot \cdot \\
A_\rho (x,\Theta) &=& A_\rho(x) + \Theta^{\mu\nu} A^{(1)}_{\mu\nu\rho}(x) + \Theta^{\mu\nu} \Theta^{\eta\sigma} A^{(2)}_{\mu\nu\eta\sigma\rho}(x) + \cdot \cdot \cdot
\eea
and when the field theory is expanded in terms of this power series Eq. (\ref{swps}) one ends up with an infinite tower of higher dimensional operators which renders the theory nonrenormalizable. However, the advantage is that this construction can be applied to any gauge theory with arbitrary matter representation. In the WM 
approach the group closure property is only found to hold for the $U(N)$ gauge theories and the matter content are found to be in the (anti)-fundamental and adjoint representations. 
Using the SW-map, Calmet \etal \cite{Calmet} first constructed a model with noncommutative gauge invariance which was close to the usual commuting  Standard Model(CSM) and is known as the {\it minimal} NCSM(mNCSM) in which they listed several Feynman rules comprising NC interaction. Intense phenomenological searches \cite{Hewett01} have been made to unravel several interesting features of this mNCSM. Hewett \etal explored several processes e.g. 
$e^+ e^- \to e^+ e^-$ (Bhabha), $e^- e^- \to e^- e^-$ (M\"{o}ller), 
$e^- \g \to e^- \g$, $e^+ e^- \to \g \g$ (pair annihilation), $\g \g \to e^+ e^-$ and $\g \g \to \g \g$ in context of NCQED. 
Recently, in a work \cite{pdas} one of us has investigated the impact of $Z$ and photon exchange in the Bhabha and the M\"{o}ller scattering and three of us have reported the impact of space-time noncommutativity in the the muon pair production at LC 
\cite{abhishodh}. Now in a generic NCQED the triple photon vertex arises to order ${\mathcal{O}}(\Theta)$, which however is absent in this minimal mNCSM. Another formulation of the NCSM came in forefront through the pioneering work by  Melic \etal \cite{Melic:2005ep}
where such a triple neutral gauge boson coupling \cite{Trampetic} appears naturally in the gauge sector. We will call this the non-minimal version of NCSM or simply NCSM. In the present work we will confine ourselves within this non-minimal version of the NCSM and use the Feynman rules for interactions given in Melic \etal \cite{Melic:2005ep}.

 In Sec. II we present the cross section of $e^+ e^- \to Z \to H H $ in the NCSM, a process which is forbidden in the CSM. The detailed numerical analysis of the pair production cross section, angular distribution and the prospects of TeV scale noncommutative geometry are presented in Sec. III. Finally, we summarize our results in Sec. IV.  
\section{Higgs pair production at future LC }
\begin{figure}[htbp]
\vspace{5pt}
\centerline{\hspace{-3.3mm}
{\epsfxsize=5cm\epsfbox{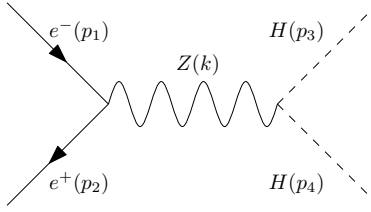}}}
\caption{Feynman diagrams for $ e^+ e^-\longrightarrow H H $ in the NCSM.}
\protect\label{feyn}
\end{figure}

As mentioned before the pair production of a neutral Higgs boson from $e^+ - e^-$ annihilation in the CSM is forbidden(at the tree level). So any excess in the predicted event rate may be interpreted as the signature of new physics. Supersymmetry and the brane world gravity are front runners ( see \cite{pkdas} and references therein). Here we explore the potential of the NCSM. In the NCSM  the pair production of a neutral Higgs boson proceeds through $ e^+ e^-\to Z \to H ~ H $ via the $s$ channel exchange of $Z$ boson. The corresponding Feynman diagram is shown in Fig. \ref{feyn}. 
The scattering amplitude using Feynman rules to order $\Theta$ for each vertex can be written as
\bea \label{unpolarized}
{i\mathcal{A}} = \frac{\pi \alpha ~ M_H^2}{\sin^2(2\theta_W) (s - m_Z^2 + i \Gamma_Z m_Z)}  \left[{\overline v}(p_2) \gamma_\mu (4 \sin^2(\theta_W) - 1 + \gamma^5)  u(p_1)\right] \nonumber \\ \times (p_3 \Theta)^\mu \left[1 + \frac{i}{2} (p_2 \Theta p_1)\right] 
\eea
where $s=(p_1 + p_2)^2$, $\alpha = e^2/4\pi$ and $\theta_W$ is the Weinberg angle. $M_H $ is the Higgs mass and $m_Z$ and $\Gamma_Z$ are the mass and decay width of the $Z$ boson. The Feynman rules required for the above result are listed in Appendix A.
The spin averaged squared-amplitude is:
\beq \label{Ampsqrd}
\overline {|{\mathcal{A}}|^2} = \frac{|{\mathcal{A}}|^2}{4} = \frac{\pi^2 \alpha^2 M_H^4}{\sin^4(2\theta_W)} \frac{[1 + (4 \sin^2\theta_W -1)^2]}{[(s-m_Z^2)^2 + \Gamma_Z^2 m_Z^2]} \left[2 (p_3 \Theta p_1) (p_3 \Theta p_2) - (p_1.p_2) [(p_3 \Theta).(p_3 \Theta)] \right]
\eeq
\noindent Explicit expressions for various dot product terms of the above equation are listed in Appendix B. The differential cross-section can be written as

\beq \label{dsigma}
\frac{d \sigma}{d \Omega} = \frac{1}{64 \pi^2 s} \frac{\lambda^{1/2}(s,M_H^2,M_H^2)}{s} {\overline {|{\mathcal{A}}|^2}} 
\eeq
where $\sigma$ = $\sigma(\sqrt{s}, \Lambda, \theta, \phi)$ and $ \lambda $ is the Kallen function defined as $\lambda(x,y,z)=x^2+y^2+z^2-2xy-2yz-2zx  $. From \ref{dsigma} we can obtain $\sigma$, $ d\sigma/d\cos\theta $ and $ d\sigma/d\phi $ as:
\bea 
\label{sigma}
\sigma &=& \int_{-1}^1 d(\cos\theta) \int_0^{2 \pi} d\phi \frac{d \sigma}{d \Omega} \\
\label{dsdcostheta}
\frac{d\sigma}{d\cos\theta} &=& \int^{2 \pi}_0 d\phi \frac{d \sigma}{d \Omega}  \\
\label{dsdphi}
\frac{d\sigma}{d\phi} &=& \int^1_{-1} d(\cos\theta) \frac{d \sigma}{d \Omega} 
\eea

\section{Numerical Analysis}
In this section, we analyze the total cross section and angular distributions of the cross section of the neutral Higgs pair production. Before making a detailed analysis, let us make some general remarks regarding the observation of non commutative effects. Since we assume 
$c_{\mu\nu}=(c_{0i},c_{ij})=(\xi_i,~\epsilon_{ijk}\chi^k)$, where $\xi_i = (\vec{E})_i$ and $\chi_k = (\vec{B})_k$ are constant vectors in a frame that is stationary with respect to fixed stars, the vectors
$(\vec{E})_i$ and $(\vec{B})_k$ point in fixed direction which are the same in all frames of reference. However, as the earth rotates around its axis and revolves around the Sun, the direction of $\vec{E}$ and $\vec{B}$ will change continuously with time dependence which is a function of the coordinates of the laboratory. The observables that are measured will thus show a characteristic time dependence. It is important to be able to measure this time dependence to verify such non commutative  theories. In our analysis, we have assumed the vectors $\vec{E}= \frac{1}{\sqrt{3}} (\hat{i} + \hat{j} + \hat{k}) $ and $\vec{B}= \frac{1}{\sqrt{3}} (\hat{i} + \hat{j} + \hat{k})$ i.e. they behave like constant vectors. This can be true only at some instant time at most.

\subsection{Pair production cross section in the NCSM}
\begin{figure}[htbp]
\vspace{-1.25in}
\centerline{\hspace{-12.3mm}
{\epsfxsize=9cm\epsfbox{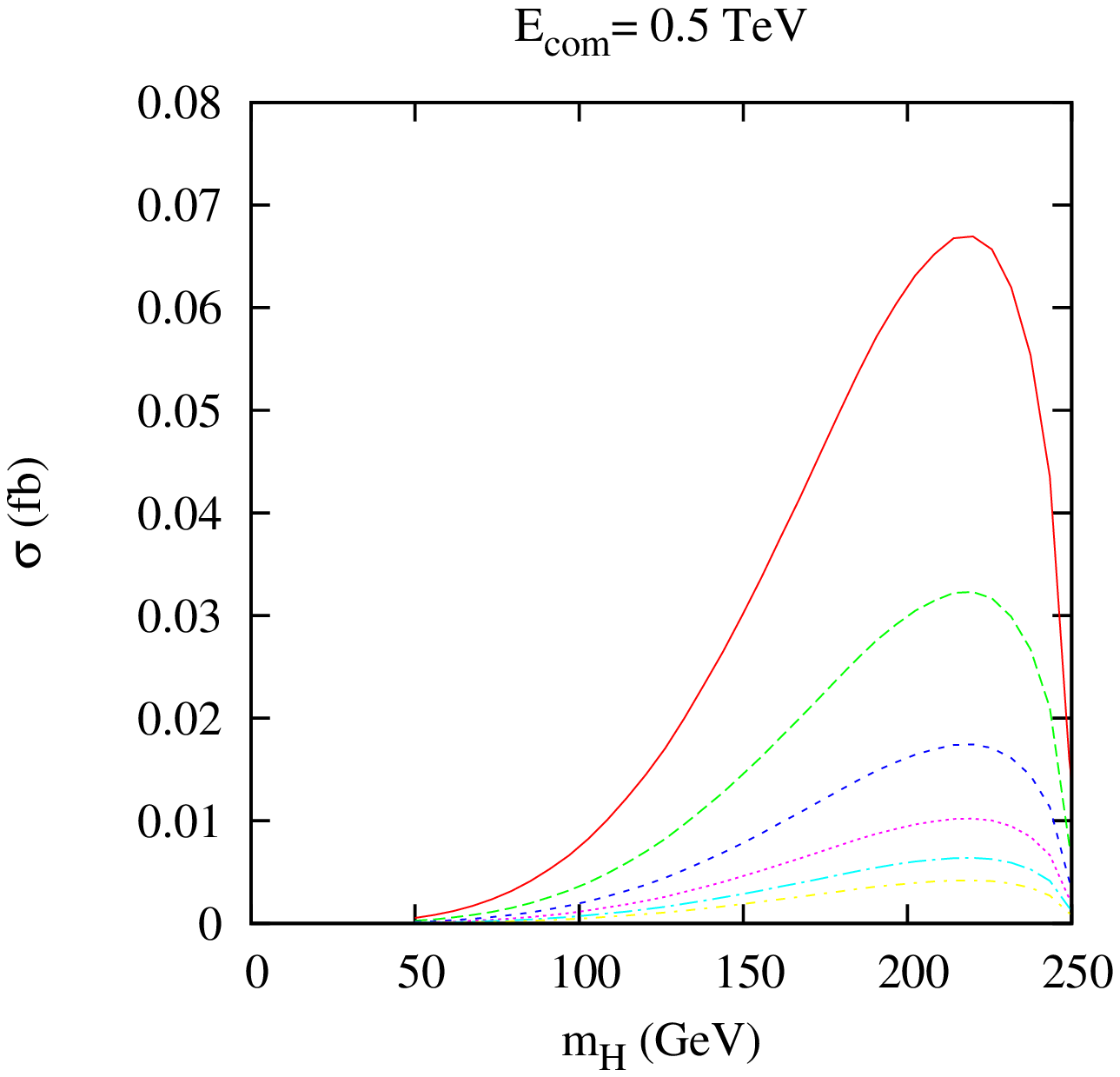}} \hspace{-0.25in} {\epsfxsize=9cm\epsfbox{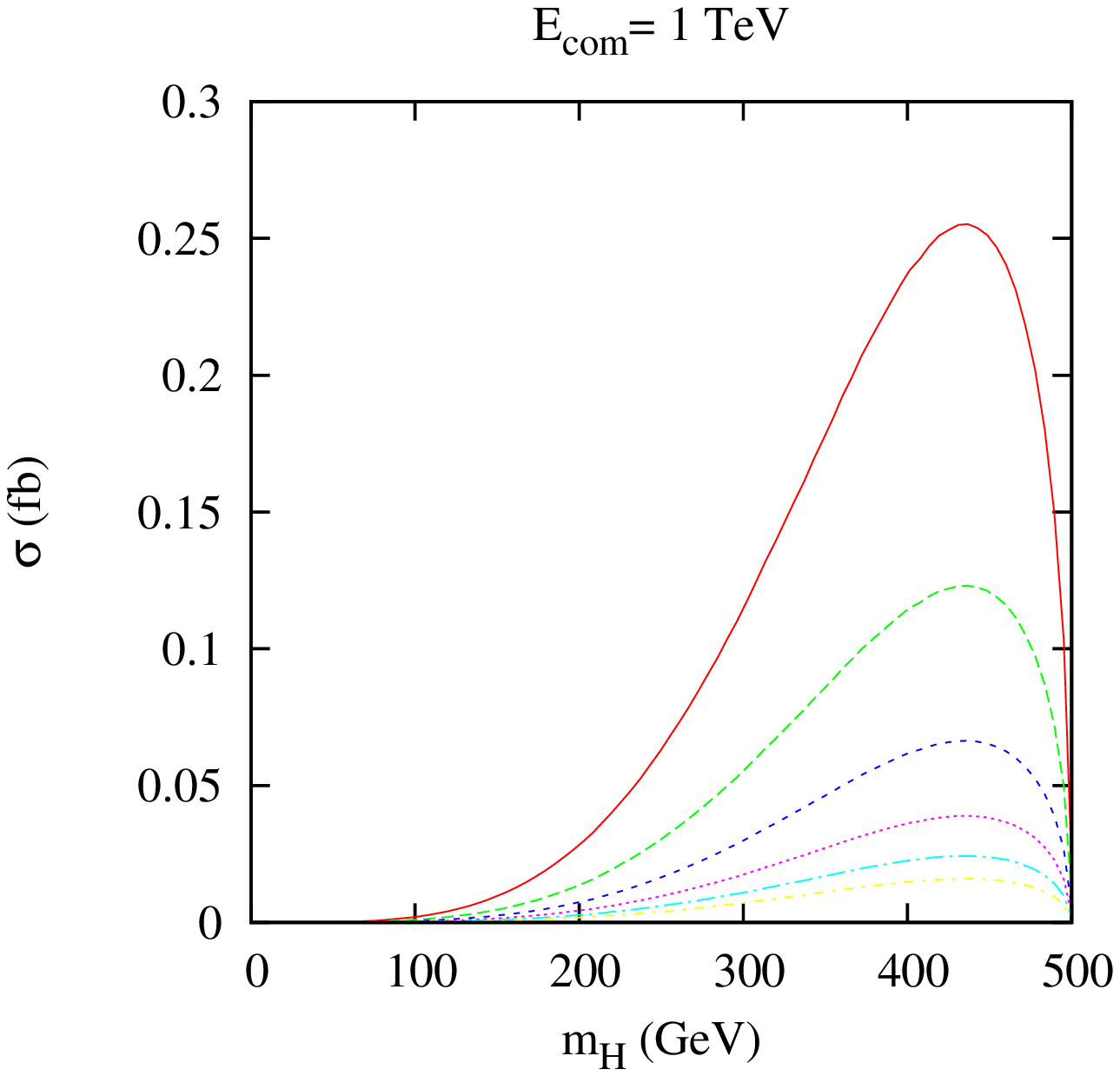}} }
\vspace*{-1.2in}
\caption{The cross section $\sigma(e^- e^+ \to Z \to H H )$ (fb) as a function of the Higgs mass $m_H$ is shown. On the l.h.s(r.h.s) figure center-of-mass energy is fixed at $\sqrt{s} =500(1000)$~GeV. While going from the top to bottom the NC scale $\Lambda$ increases from $500$ GeV to $1000$ GeV in steps of $100$ GeV.}
\protect\label{sigplot}
\end{figure}

As we have mentioned earlier the Higgs pair production is forbidden in the CSM. Any signature of such an event will correspond to new physics and NCSM is one of the promising candidates among such a class of new physics models.  In Fig. (\ref{sigplot}) we have plotted the total cross section $\sigma(e^- e^+ \to Z \to H~H)$ as a function of the Higgs mass $m_H$(GeV). The machine energy is fixed at  $E_{com}(=\sqrt{s}) = 500 $ GeV and $1000$ GeV as shown above. 
In each figure while going from the top to the bottom, the curves correspond to $\Lambda = 500,~600,~700,~800,~900$ and $1000$~GeV, respectively. Note that the pair production cross sections are maximum at $m_H = 220(437)$ corresponding to the machine energy $\sqrt{s} = 500(1000)$ GeV. Assuming an integrated luminsoity of the futute LC about $\mathcal{L} = 500~fb^{-1}$, we have predicted the number of events in the NCSM. The results, the number of events $N~$($yr^{-1}$) as a function of $\Lambda$ corresponding to the machine energy $\sqrt{s}=500$ GeV and $1000$ GeV are presented in Table 1. Note that if we increase $\Lambda$ from $500$ GeV to $1000$ GeV, the number of events(NC signals) $N (yr^{-1})$ decreases from $34(128)$ per year to  $2(8)$ per year. 
So a maximum of $34$ and $128$ events (NC signal) per year are expected to be observed at the Future LC.
\newpage

\vspace*{-0.2in} 
\begin{center}
Table 1
\end{center}
\vspace*{-0.15in}
\begin{center}
\begin{tabular}{|c|c|c|c|c|c|c|c|c|c|c|}
\hline
$\sqrt{s}$ & $\Lambda$ & $\sigma$(fb) & ${\mathcal{L}}(fb^{-1})$  & N ($yr^{-1}$)&& $\sqrt{s}$ & $\Lambda$ & $\sigma$(fb) & ${\mathcal{L}}(fb^{-1})$  & N ($yr^{-1}$\\
\hline
\hline
   & 500 &0.0670 & 500 & 34 &&&500 &0.2552 & 500& 128\\
\hline
  & 600 & 0.0320 & 500 & 16 &&&600&0.1231 &500& 62\\
\hline
  500 & 700 & 0.0170 & 500 &9 &&1000&700& 0.0664&500& 33\\
\hline
  & 800 & 0.0100& 500 & 5 &&&800& 0.0389&500& 19\\
\hline
  & 900 & 0.0064 & 500 & 3 &&&900& 0.0243&500& 12\\
\hline
  & 1000 & 0.0042 & 500 & 2 &&&1000& 0.0159&500& 8\\
\hline
\end{tabular}
\end{center}
\noindent {\it Table 1: The yearly number of events(NC signals) with the increase in the NC scale $\Lambda$ is shown. The integrated luminosity of the LC is assumed to be ${\mathcal{L}}=500~fb^{-1}$. }.

 \noindent  These are to be compared with the zero event prediction in the CSM where this process is forbidden. 
\subsection{Angular distribution of muon pair production $e^- e^+ \to Z \to H H $ in the NCSM }

\begin{figure}[htbp]
\label{sigs}
\vspace{-1.2in}
\centerline{\hspace{-12.3mm}
{\epsfxsize=9cm\epsfbox{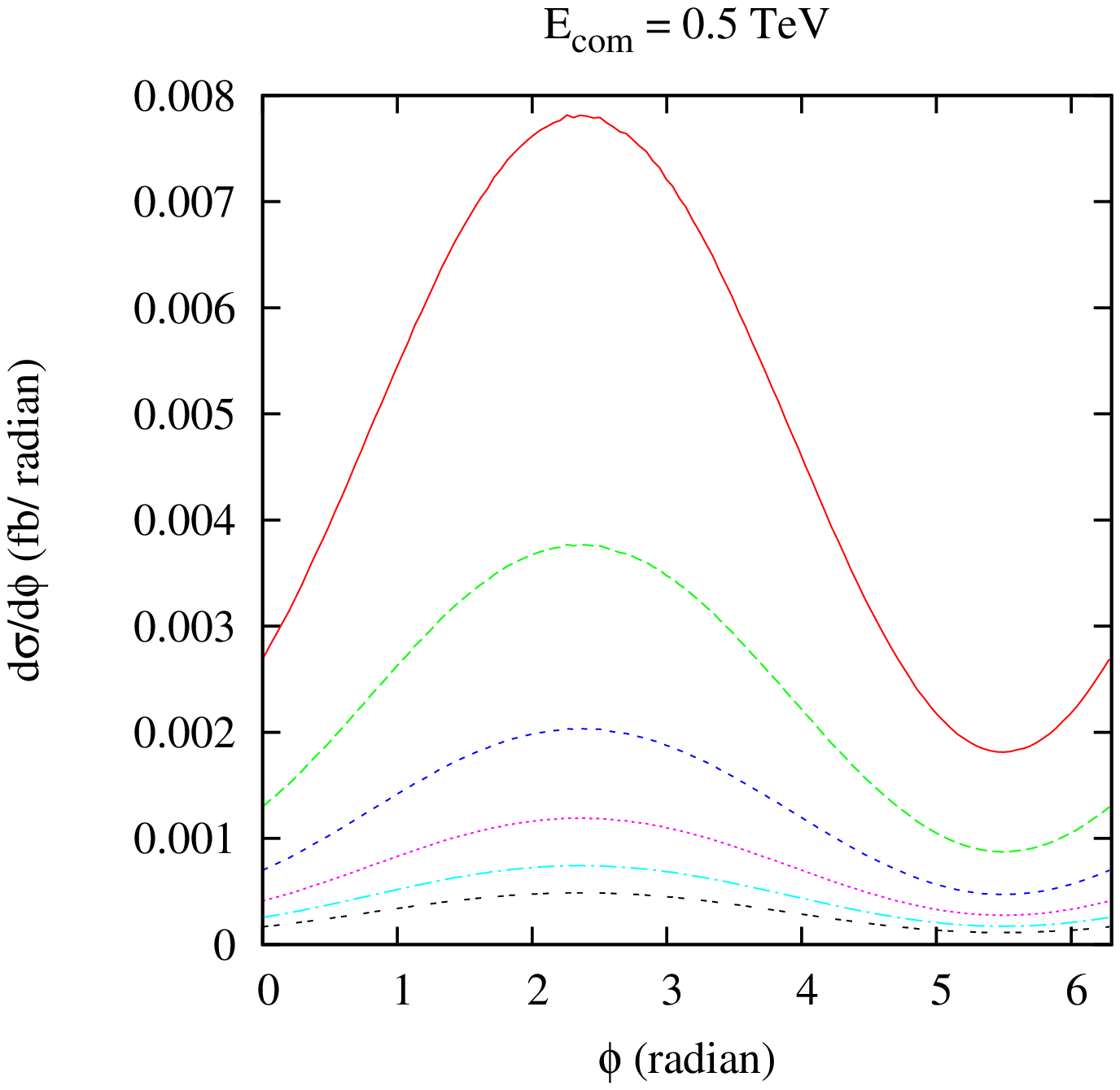}} \hspace{-0.25in} {\epsfxsize=9cm\epsfbox{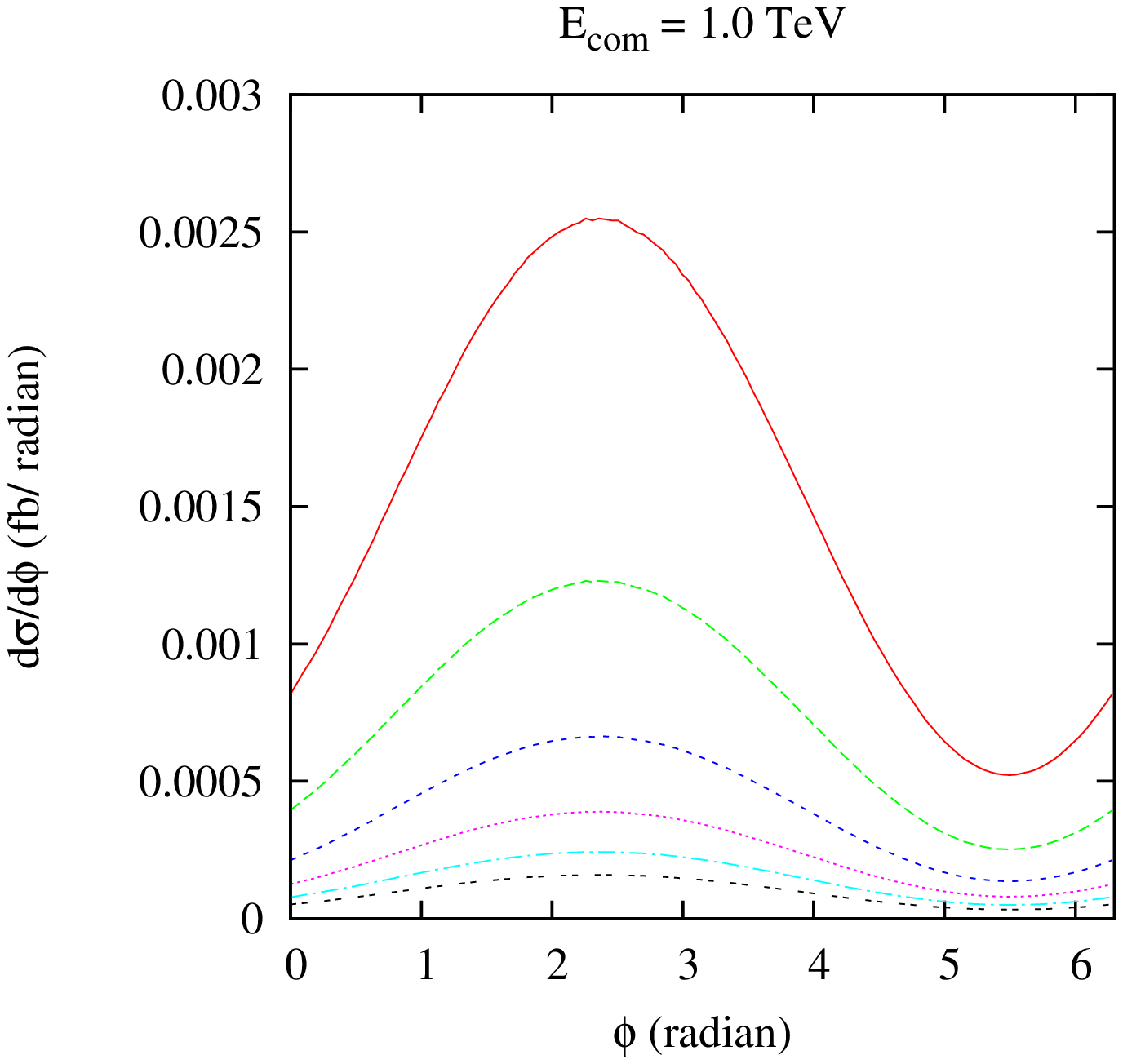}} }
\vspace*{-1.15in}
\caption{{The $ \frac{d\sigma}{d\phi} $($fb/rad$) distribution as a function of $\phi$(in rad) is shown. The lowest horizontal curve at zero is due to the CSM, whereas the plots above the horizontal one, as we move up correspond to $\Lambda = 1.0, 0.9, 0.8, 0.7, 0.6 $ and $0.5$ TeV, respectively in the NCSM. }}
\protect\label{dsdphiplot}
\end{figure}

{\it {Azimuthal distribution}}: The angular distribution of the final state particles is a useful tool to understand the nature of new physics. Next we will see how the azimuthal distribution can be used to seperate out the noncommutative geometry(and thus the NCSM) from the other type of new physics models e.g supersymmetry, brane world gravity, unparticle scenario, little Higgs models etc. In Fig. \ref{dsdphiplot} we show $\frac{d\sigma}{d\phi}$ as a function of the azimuthal angle 
$\phi$.  
For the angular analysis study, we consider two cases corresponding to the c.o.m energies 
$\sqrt{s} = 500$ GeV and $1000$ GeV, respectively. Corresponding to a particular c.o.m energy we vary the NC scale $\Lambda$ from $0.5$ TeV to $1.0$ TeV. The distribution $d\sigma/d\phi$ is completely flat at zero in the CSM, since the process is forbidden there. Other plots above the horizontal zero line, as we move up correspond to $\Lambda = 1.0, 0.9, 0.8, 0.7, 0.6 $ and $0.5$ TeV, respectively in the NCSM.  The departure from the flat behaviour is due to $ p_3 \Theta p_1$,  $p_3 \Theta p_2$ and $(p_3 \Theta).(p_3 \Theta)$ terms in Eqns. \ref{Ampsqrd} that bring in the $\phi$ dependence which is observed in Fig. \ref{dsdphiplot}. Interestingly, in each of the two figures the curves shows maxima and minima. The maxima arises at $\phi = 3 \pi/4$ rad, whereas the minima is found to be located at $\phi = 7 \pi/4$ rad. Note that if we set $\Lambda = \infty$, the lowest(zero) horizontal curve in the CSM is recovered.  
 
 Note that such an azimuthal distribution clearly reflects the exclusive nature of the noncommutative geometry which is rarely to be found in other classes of new physics models. 

\begin{figure}[htbp]
\label{sigs}
\vspace{-1.2in}
\centerline{\hspace{-12.3mm}
{\epsfxsize=9cm\epsfbox{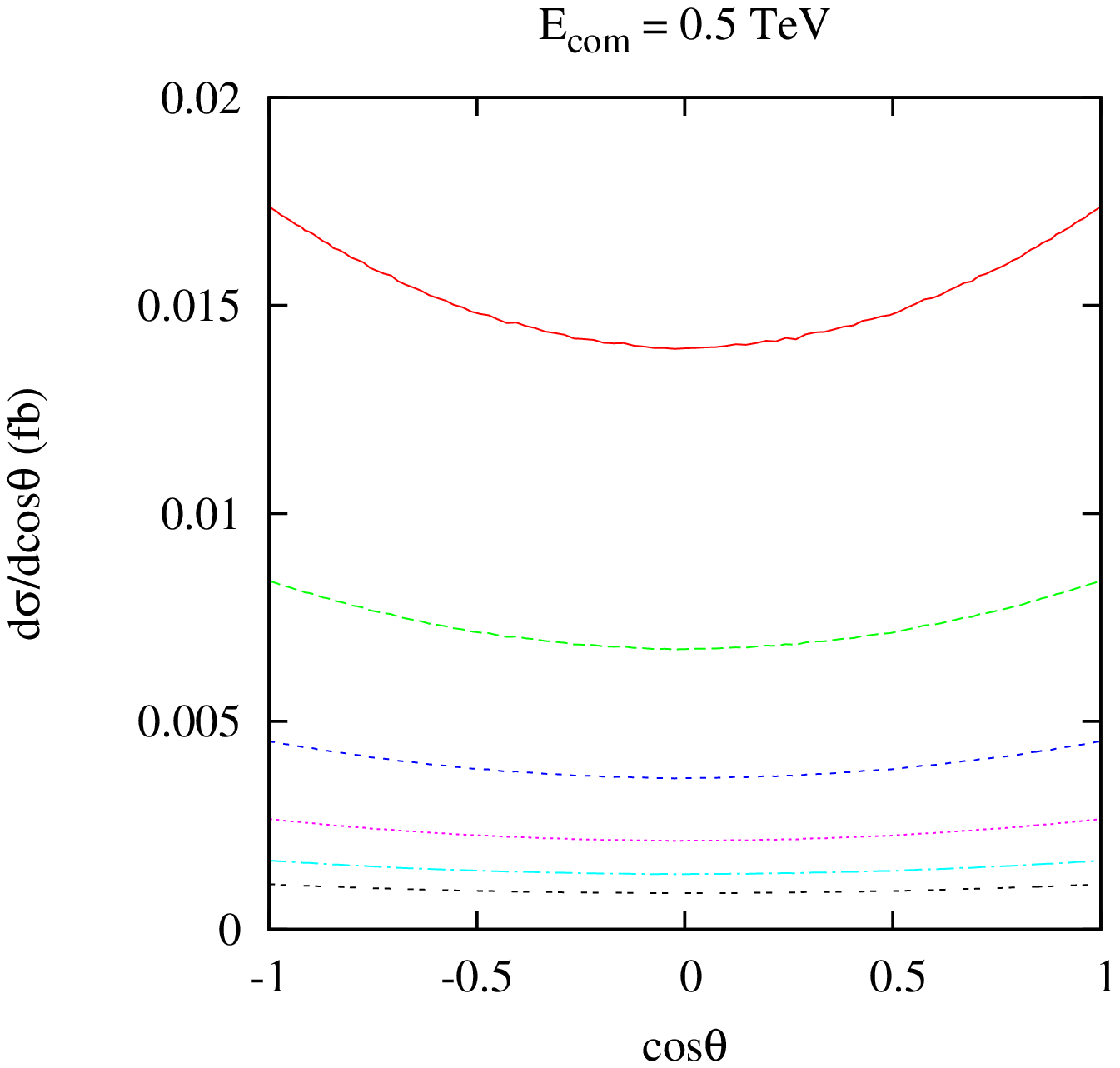}} \hspace{-0.25in} {\epsfxsize=9cm\epsfbox{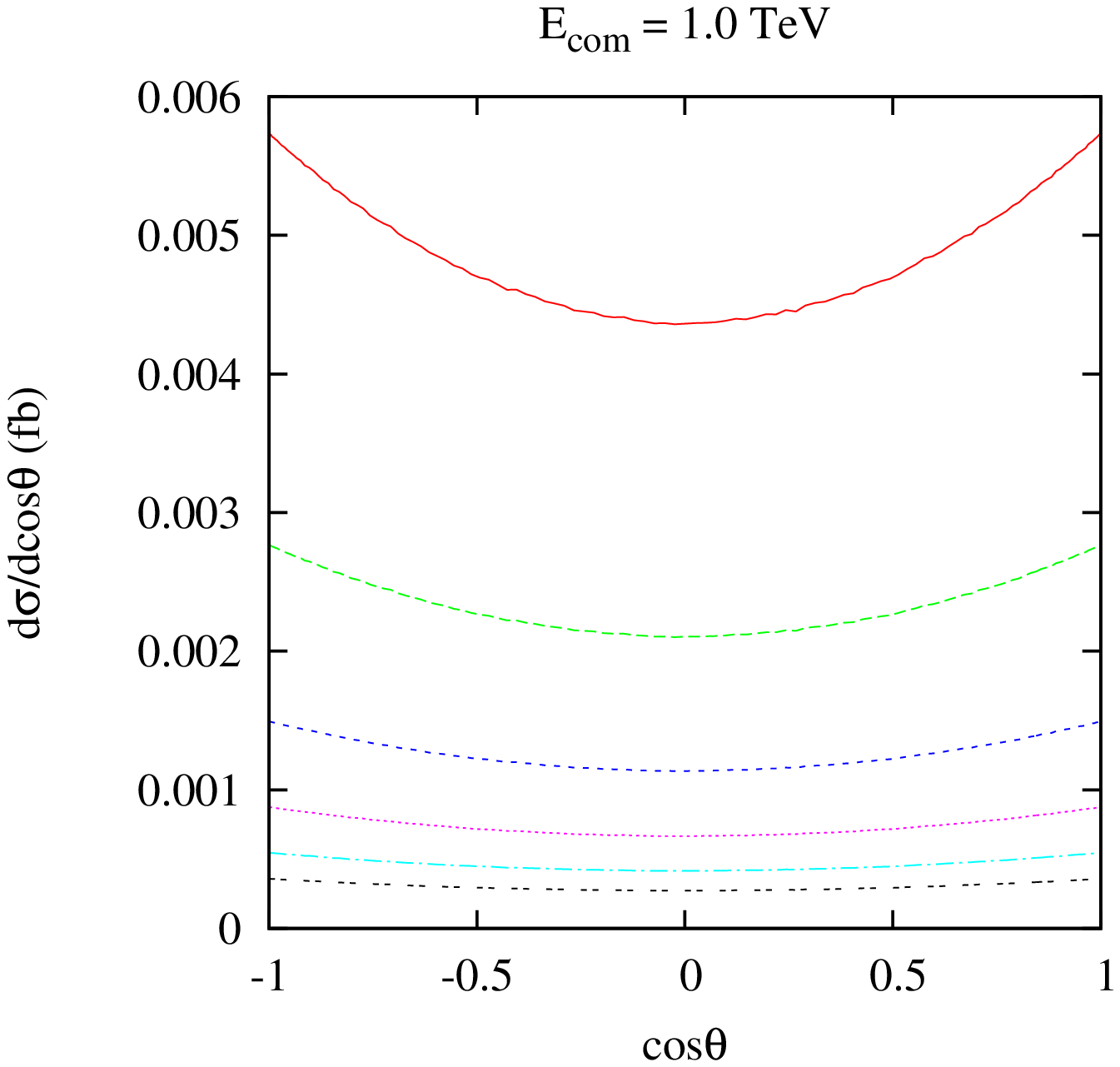}} }
\vspace*{-1.15in}
\caption{{The $ \frac{d\sigma}{dcos\theta} $($fb$) distribution as a function of $cos\theta$(in rad) is shown. The lowest horizontal curves(at zero value) in both figures are due to the CSM, whereas the plots above the horizontal one, as we move up correspond to $\Lambda = 1.0, 0.9, 0.8, 0.7, 0.6 $ and $0.5$ TeV, respectively in the NCSM. }}
\protect\label{dsdcosthetaplot}
\end{figure}

{\it Polar distribution: } To extract the exclusive nature of noncommutative geometry, like azimuthal distribution, the polar distribution might also be useful. In Fig. \ref{dsdcosthetaplot}, $\frac{d\sigma}{dcos\theta}$ is plotted as a function of $cos\theta$. 
Note that the distribution around the $cos\theta = 0$ line of Fig. \ref{dsdcosthetaplot} is completely symmetric.  The lowermost plot(zero horizontal curve) in Fig. \ref{dsdcosthetaplot} corresponds to the polar distribution in the CSM(reflecting the fact that the process is forbidden in the CSM) and the plots, as we move up correspond to $\Lambda = 1.0, 0.9, 0.8, 0.7, 0.6 $ and $0.5$ TeV, respectively, gets more and more curved, which is a unique features of the NCSM. The uppermost curve in the figure corresponding to $\Lambda = 0.5$ TeV exhibits maximal deviation from the lowermost CSM curve (obtained by setting $\Lambda \to \infty$).  

\subsection{Constraining the NC scale $\Lambda$ using experimental bound on Higgs mass $m_H$ }
We found in an earlier subsection that the pair production cross section is maximum at $m_H = 220$ GeV and $437$ GeV corresponding to the c.o.m energy $\sqrt{s} = 500$ GeV and $1000$ GeV. The corresponding NC scale is $\Lambda = 500$. In Fig. 5 we make some contour plots in the $m_H - \Lambda$ plane corresponding to event production rate per year $N (yr^{-1})$ with 
\begin{itemize}
\item Scenario I : $2 \le N  \le 34$ and the c.o.m energy $\sqrt{s} = 500$ GeV,
\item Scenario II: $8 \le N  \le 128$ and the c.o.m energy $\sqrt{s} = 1000$ GeV.
\end{itemize} 
\begin{figure}[htbp]
\centerline{\hspace{-12.3mm}
{\epsfxsize=10cm\epsfbox{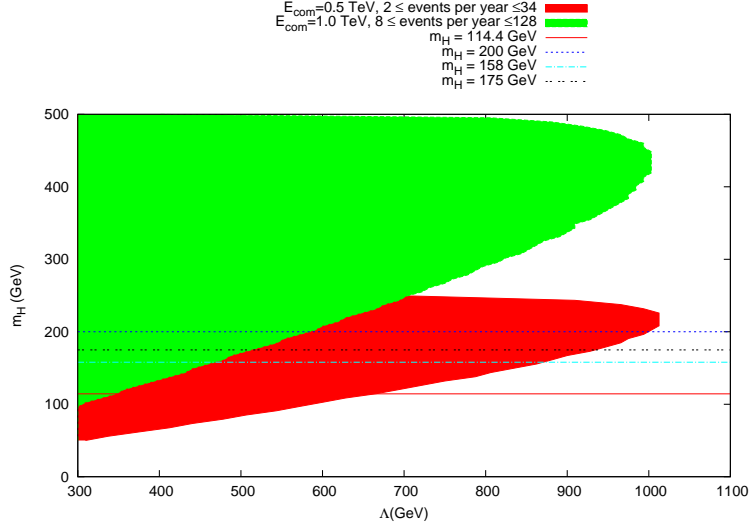}} }
\vspace*{-0.15in}
\caption{{The contour plot in the $m_H - \Lambda$ plane obtained by setting $2 \le N(yr^{-1}) \le 34$(lower) and $8 \le N(yr^{-1}) \le 128$(upper) corresponding to $\sqrt{s} = 500$ GeV and $1000$ GeV, respectively.  The lowermost and uppermost horizontal lines correspond to the lower and upper bound on the Higgs mass $m_H$ which follows from the LEP II direct search of Higgs boson and the global electro-weak fit. The combined CDF and DO data at Tevatron excludes $158$ GeV $\le m_H \le 175$ GeV. Using these we obtain bounds on the bound NC scale $\Lambda$. (See the text for further details). }}
\protect\label{contourplot}
\end{figure}
The following results are in the order:
\begin{itemize}
\item The direct search of Higgs boson at LEP II gives a lower bound on Higgs mass $m_H$ as $114.4$ GeV. Incorporating this in 
Fig (\ref{contourplot}), one finds : (i)$\Lambda > 666$ GeV in scenario I and (ii) $\Lambda > 347$ GeV in scenario II. 
\item The combined CDF and DO data at Tevatron, Fermilab excludes $m_H$ lying between $158 $GeV $\le m_H \le 175$ GeV. Translating this in Fig (\ref{contourplot}), one finds  $\Lambda \le 872$ GeV and $\Lambda \ge 938$ GeV in scenario I and 
$\Lambda \le 468$ GeV and  $\Lambda \ge 515$ GeV in scenario II.
\item The Global Electro-weak fit suggests $m_H \le 200$ GeV. Translating this in Fig (\ref{contourplot}), one finds 
$\Lambda \le 999$ GeV in scenario I and $\Lambda \le 581$ GeV in scenario II.
\end{itemize}

Altogether, the direct searches of Higgs at LEP II and Tevatron and global electroweak fit to $m_H$ gives rise the following bound on $m_H$: 
\begin{itemize}
\item $665$ GeV $ \le \Lambda \le 872$ GeV and  $938$ GeV $ \le \Lambda \le 998$ GeV  in scenario I.
\item $347$ GeV $ \le \Lambda \le 468$ GeV and  $515$ GeV $ \le \Lambda \le 581$ GeV  in scenario II. 
\end{itemize}

\section{Conclusion}
The idea of the TeV scale space-time noncommutativity draws a lot of attention in the physics community. We explored the impact of spacetime noncommutivity in the Higgs pair production $e^+ e^- \to Z \to  H H $ within the NCSM. The plots showing the cross section $\sigma(e^+ e^- \to Z \to  H H )$ as a function of Higgs mass $m_H$ at a fixed machine energy $E_{com}$,  suggests that at a particular $m_H$ the pair production cross section is maximum. Corresponding to the macine energy $\sqrt{s} = 500$ GeV and $1000$ GeV, the peak(where the cross section and thus the event rate is maximum) is located at $m_H = 220$ GeV and $m_H = 437$ GeV. The maximum number of events corresponding to machine energy $\sqrt{s} = 500$ GeV and $1000$ GeV turn out to be $34~(yr^{-1})$ and $128~(yr^{-1})$, respectively with no events of Higgs pair production in the CSM. Another interesting feature of our study is the angular distribution. The azimuthal distribution $d\sigma/d\phi$ which is supposed to be completely $\phi$ symmetric(zero in the CSM), deviates substantially from its flat(zero) behaviour in the NCSM. The deviation sees an enhancement with the lowering of NC scale $\Lambda$. Note that such a non-trivial azimuthal distribution is a unique feature of the noncommutative standard model and it is rarely found in any class of new physics models(e.g.supersymmetry, Randall-Sundrum model etc) which supports neutral Higgs pair production. Finally, we study $d\sigma/dcos\theta$ as a function of $cos\theta$: zero in the CSM deviates from it's CSM behaviour and develops an asymmetry around the $cos\theta = 0$ curve in the NCSM. There is a clear departure of the distribution obtained in the NCSM from the one obtained in the CSM. 
We make the contour plots in the plane of $m_H - \Lambda$ corresponding to the event rate $2 \le N(yr^{-1}) \le 34$ and $8 \le N(yr^{-1}) \le 128$. Finally the direct searches of Higgs at LEP II and Tevatron and global electroweak fit to $m_H$ gives rise the following bound on $m_H$: (i) $665$ GeV $ \le \Lambda \le 872$ GeV and  $938$ GeV $ \le \Lambda \le 998$ GeV  in scenario I and (ii) $347$ GeV $ \le \Lambda \le 468$ GeV and  $515$ GeV $ \le \Lambda \le 581$ GeV  in scenario II. 
 So, the noncommutative geometry is found to be quite rich in terms of its phenomenological implications and it is worthwhile to explore several other processes  which might be interesting and potentially relevant for the Linear Collider experiments. 
\begin{acknowledgments}
The work of P.K.Das is supported by the DST Fast Track Young Scientist project SR/FTP/PS-11/2006. P. K. Das would also like to thank Dr. Chandradew Sharma and Dr. Tarun Kumar Jha of the Physics department of BITS-Goa for several useful discussions that he had with them.   
\end{acknowledgments}
\appendix

\section{Feynman rules to order ${\mathcal{O}}(\Theta)$ }
The Feynman rule for the vertex $f(p_{in})- f(p_{out}) - Z(k)$ is \cite{Melic:2005ep}
\bea
\frac{e}{sin2\theta_W} \left[i \gamma_\mu \Gamma_A^- \right] + \frac{e}{2 sin2\theta_W} \left[(p_{out} \Theta p_{in}) \gamma_\mu \Gamma_A^- 
 - (p_{out} \Theta)_\mu \Gamma_A^+ (\pinsla -  m_f) - (\poutsla -  m_f) \Gamma_A^- (\Theta p_{in})_\mu \right]. \nonumber\\
\eea
\noindent The momentum conservation reads as $p_{in} + k = p_{out}$. Similarly, the Feynman rule for the interaction vertex vertex $H(p)-H(p)-Z(k)$ is:
\beq
\frac{g~ M_H^2~ (p \Theta)_\mu}{4 \cos\theta_W}.
\eeq

\noindent Here $(p\Theta)_\mu = p_{\mu \nu} \Theta^\nu $,~ 
$\Gamma_A^\pm = (c_V^e \pm c_A^e \gamma_5)$ . Also $ p_{out} \Theta p_{in} = p_{out}^\mu  \Theta_{\mu \nu}  p_{in}^\nu = -p_{in} \Theta p_{out}$.

\section{Momentum prescriptions and dot products}
Working in the center of momentum frame and ignoring the electron mass, we can specify the 4 momenta of the particles as follows:
\bea
\label{prescstart} p_1 &=& \left(\frac{\sqrt{s}}{2}, 0, 0, \frac{\sqrt{s}}{2}\right)\\
p_2 &=& \left(\frac{\sqrt{s}}{2}, 0, 0, -\frac{\sqrt{s}}{2}\right)\\
p_3 &=& \left(\frac{\sqrt{s}}{2},k' \sin\theta \cos\phi, k' \sin\theta \sin\phi, k' \cos\theta \right)   \\
p_4 &=& \left(\frac{\sqrt{s}}{2},-k' \sin\theta \cos\phi,-k' \sin\theta \sin\phi, -k' \cos\theta \right), \\
k' &=& \frac{\sqrt{s}}{2} \sqrt{1- \frac{4M_H^2}{s} } \nonumber
\eea
where  $\theta$ is the scattering angle made by the $3$-momentum vector $p_3$ of $H(p_3)$ with the +ve Z axis and $\phi$ is the azimuthal angle. 
We note that the antisymmetric $\Theta_{\mu \nu}$ has $6$ independent components corresponding to $c_{\mu \nu} = (c_{0i}, c_{ij})$ with $i,j=1,2,3$. Assuming all of them are non-vanishing they can be written in the form
\bea
c_{0i} &=& \frac{\xi_i}{\Lambda^2}, \\
\label{prescend}
c_{ij} &=& \frac{\epsilon_{ijk} \chi^k}{\Lambda^2}.
\eea
The antisymmetric $\Theta_{\mu \nu}$ is analogous to the field tensor $F_{\mu \nu}$ where $\xi_i$ and $\chi_i$ are like the components of the Electric and Magnetic Field vectors.
Setting $\xi_i=(\vec{E})_i = \frac{1}{\sqrt{3}},~i=1,2,3$ and $\chi_i= (\vec{B})_i = \frac{1}{\sqrt{3}},~i=1,2,3$( noting the fact that $\chi_i = - \chi^i$, $\xi_i = - \xi^i$ and $\xi_i \xi^j = \frac{1}{3} \delta_i^j$ and $\chi_i \chi^j = \frac{1}{3} \delta_i^j$, we find
\bea
p_1.p_2 &=& \frac{s}{2},\\
p_2 \Theta p_1 &=&  \frac{s}{2 \sqrt{3} \Lambda^2}, \\
p_3 \Theta p_1 &=&  \frac{-s}{4 \sqrt{3} \Lambda^2} \left[\sqrt{1-\frac{4M_H^2}{s}} (2 \sin\theta \sin\phi + \cos\theta) -1\right],\\
p_3 \Theta p_2 &=&  \frac{-s}{4 \sqrt{3} \Lambda^2} \left[\sqrt{1-\frac{4M_H^2}{s}} (2 \sin\theta \cos\phi + \cos\theta) +1\right],\\
(p_3\Theta)_0 &=&  \frac{-\sqrt{s}}{2 \sqrt{3} \Lambda^2} \left[\sqrt{1-\frac{4M_H^2}{s}} (\sin\theta \cos\phi + \sin\theta \sin\phi +\cos\theta )\right],\\
(p_3\Theta)_1 &=& \frac{\sqrt{s}}{2 \sqrt{3} \Lambda^2} \left[1 - \sqrt{1-\frac{4M_H^2}{s}} (\cos\theta - \sin\theta \sin\phi ) \right],\\
(p_3\Theta)_2 &=& \frac{\sqrt{s}}{2 \sqrt{3} \Lambda^2} \left[1 - \sqrt{1-\frac{4M_H^2}{s}} (sin\theta \cos\phi-\cos\theta) \right],\\
(p_3\Theta)_3 &=& \frac{\sqrt{s}}{2 \sqrt{3} \Lambda^2} \left[1 + \sqrt{1-\frac{4M_H^2}{s}}( sin\theta \cos\phi-\sin\theta \sin\phi) \right].
\eea



\end{document}